\documentclass[10pt]{article}
\pdfoutput=1
\usepackage{amssymb,amsmath,mathrsfs,enumerate}
\usepackage{graphicx,rotate,multicol}
\usepackage[margin=10pt,labelfont=bf,font={small,it}]{caption}
\usepackage{cite}
\usepackage{epsfig}
\usepackage[colorlinks=true,
            linkcolor=red,
            urlcolor=blue,
            citecolor=blue]{hyperref}
\usepackage{array}
\usepackage{color}
\long\def\rpl#1!!#2!!{\textcolor{red}{#1} \textcolor{blue}{#2}}

\long\def\rpl#1!!#2!!{\textcolor{red}{#1} \textcolor{blue}{#2}}

\def \order(#1){{\cal O} \left(#1 \right)}

\evensidemargin=\oddsidemargin
\def\Eqn#1{Eq.\ (\ref{#1})}

\allowdisplaybreaks
\newcommand{\bea}{\begin{eqnarray}}
\newcommand{\beq}{\begin{equation}}
\newcommand{\eea}{\end{eqnarray}}
\newcommand{\eeq}{\end{equation}}
\newcommand{\LS}{\Lambda_{\rm SUSY}}
\newcommand{\Lf}{\Lambda_f}

\newcommand{\nn}{\nonumber}

\newcommand{\lsim}{\raise0.3ex\hbox{$\;<$\kern-0.75em\raise-1.1ex\hbox{$\sim\;$}}}
\newcommand{\gsim}{\raise0.3ex\hbox{$\;>$\kern-0.75em\raise-1.1ex\hbox{$\sim\;$}}}
\newcommand{\eq}[1]{Eq.~(\ref{#1})}
\newcommand{\unity}{{\hbox{1\kern-.8mm l}}}

\begin{document}

\hfill {\small FTUV-16-0719, IFIC-16-51}\\[0.5cm]

\begin{center}
	{\Large \bf Effective Theories of Flavor and the Non-Universal MSSM} \\
	\vspace*{1cm} {\sf Dipankar	Das\footnote{dipankar.das@uv.es}, ~M. L. L\'opez-Ib\'a\~nez\footnote{m.luisa.lopez-ibanez@uv.es}, ~M. Jay P\'erez\footnote{mjperez@ific.uv.es}, ~Oscar Vives\footnote{oscar.vives@uv.es}} \\
	\vspace{10pt} {\small\em Departament de F\'{i}sica T\`{e}orica, Universitat de Val\`{e}ncia and IFIC, Universitat de Val\`{e}ncia-CSIC  \\
	Dr. Moliner 50, E-46100 Burjassot (Val\`{e}ncia), Spain}
	
	\normalsize
\end{center}

\begin{abstract}
Flavor symmetries \`{a} la Froggatt-Nielsen~(FN) provide a compelling way to explain the hierarchies of fermionic masses and mixing angles in the Yukawa sector. In  Supersymmetric~(SUSY) extensions of the Standard Model where the mediation of SUSY breaking occurs at scales larger than the breaking of flavor, this symmetry must be respected not only by the Yukawas of the superpotential, but by the soft-breaking masses and trilinear terms as well. In this work we show that contrary to naive expectations, even starting with completely flavor blind soft-breaking in the full theory at high scales, the low-energy sfermion mass matrices and trilinear terms of the effective theory, obtained upon integrating out the heavy mediator fields, are strongly non-universal. We explore the phenomenology of these SUSY flavor models after the latest LHC searches for new physics.
\end{abstract}

\bigskip
\section{Introduction}
\label{s:intoduction}
Since the discovery of the muon in cosmic rays at the beginning of the 20th century, the flavor puzzle remains one of the biggest open questions in high-energy physics. This puzzle derives from the bizarre flavor structures present in the Standard Model (SM) and the mystery, if any, behind their origin. Although the SM is able to accommodate all  known flavor parameters in its Yukawa matrices, the values of these parameters are completely arbitrary and can only be fixed from experimental measurements.

Using symmetries to interpret a large number of apparently arbitrary parameters is a strategy that has been remarkably successful for similar problems in the past. In this spirit, the FN mechanism\cite{Froggatt:1978nt} uses a spontaneously-broken flavor symmetry to explain the large hierarchies observed in the fermionic masses and mixing angles. The SM Yukawas are  understood as effective couplings in powers of a small parameter $\varepsilon \equiv \langle \phi\rangle/\Lambda_f$, where $\langle\phi\rangle$ represents the vacuum expectation value~(vev) of the flavon field, $\phi$, and $\Lambda_f$ represents the characteristic mass scale of the flavor mediators.  The different elements of the Yukawa matrices can then be explained by suitable assignments of flavor charges to  different fields. However, there are a plethora of possible choices for the flavor symmetry and its breaking which are consistent with the existing flavor data\cite{Babu:2009fd,Raidal:2008jk}: Abelian or non-Abelian, continuous or discrete and with a variety of gauge groups and representations\cite{Leurer:1992wg,Leurer:1993gy,Nir:1993mx,Dine:1993np,Ibanez:1994ig,Barbieri:1995uv,Pomarol:1995xc,Dudas:1995yu,Nir:1996am,Binetruy:1996cs,Dudas:1996fe,Binetruy:1996xk,Barbieri:1996ww,Irges:1998ax,Choi:1998wc,Barbieri:1999km,Ma:2001dn,King:2001uz,Nir:2002ah,Babu:2002dz,Ross:2002fb,Altarelli:2002sg,King:2003rf,Dreiner:2003yr,Ross:2004qn,Chankowski:2005qp,deMedeirosVarzielas:2005qg,Altarelli:2005yx,deMedeirosVarzielas:2006fc,Luhn:2007sy,Ishimori:2009ew,Ishimori:2010au,Altarelli:2010gt,Babu:2011mv,King:2013eh,Chen:2014wiw,Kile:2013gla}.

In this paper we choose to focus on the FN mechanism as the solution to the flavor puzzle.
Although perhaps the best way to test this idea would be to directly produce the flavor mediators,  in practice this is impossible if the flavor symmetry is broken at energies much higher than the electroweak~(EW) scale. Given the fact that the Yukawa couplings are dimensionless and depend only on the ratio $\varepsilon = \langle \phi\rangle/\Lambda_f$, in principle the flavor symmetry could be broken at any scale above the EW scale, even close to $M_{\rm Pl}$.  Therefore, if the Yukawa couplings are the only remnant of flavor symmetry breaking, we may never be able to unravel the origin of flavor.

Nevertheless, as the LHC continues to explore the TeV scale \cite{Aad:2016jxj,Aad:2016qqk,Aad:2016tuk,Aad:2016eki,Aaboud:2016tnv,Aaboud:2016zdn,Aaboud:2016lwz,Khachatryan:2016kdk,Khachatryan:2016xvy,Khachatryan:2016kod,Khachatryan:2016uwr,Khachatryan:2016fll}, we can hope that new physics associated with the breaking of the flavor
symmetry may emerge in this energy region. In many extensions of the SM proposed in the literature, this new
physics is not completely flavor-blind, and the new interactions have a non-trivial flavor dependence. If this is indeed the case, we expect to
find new signatures of flavor at the LHC, complementary to the information already present in the Standard Model.  

Supersymmetric extensions of the Standard Model provide a nice example of
this point\cite{Nilles:1983ge,Haber:1984rc,Martin:1997ns,Chung:2003fi}. In the SM, we can only measure the fermion masses and the left-handed mixings (the relative misalignment between $Y_u$ and $Y_d$ in the quark sector) in the Yukawa matrices; right-handed mixings are completely unobservable
as each right-handed matter field couples to a single flavor
matrix that can always be diagonalized. However, in SUSY extensions of the SM
right-handed fields couple also through their trilinear interactions and moreover may have
non-trivial soft-mass matrices. In general these three flavor structures can not be simultaneously diagonalized and their relative misalignment becomes observable\cite{Botella:2004ks}, providing  a way to explore whether
the mixings in the right-handed sector are small (analogous to the
case of the CKM left-handed mixings) or large (similar perhaps to the
PMNS neutrino mixings). Similarly, the measure of left-handed mixings
in the leptonic fields can clarify whether the large PMNS mixings are
due to large Yukawa mixings or to the Majorana nature of the neutrino
masses through a type-I seesaw mechanism. 

Indeed, if SUSY is found at the LHC and these new flavor structures are observed,
they may help to elucidate the origin of flavor. 
If a flavor symmetry \`a la Froggatt-Nielsen\cite{Froggatt:1978nt} is responsible for the observed Yukawa couplings in a SUSY theory, this flavor symmetry should apply at the level of the superfields of the theory, which include both the SM fermions and their superpartners.
After the flavor symmetry is broken spontaneously, we will obtain the Yukawa
couplings as well as non-trivial corrections to the soft-breaking masses and trilinear couplings\cite{Raidal:2008jk,Ramage:2003pf,Calibbi:2008qt,Calibbi:2009ja,Lalak:2010bk,Calibbi:2011dn,Babu:2014sga}.

Theoretical and phenomenological constraints  force SUSY breaking to take
place in a hidden sector with no renormalizable couplings to the observable
fields. Then the soft SUSY-breaking terms may arise from the non-renormalizable
operators in the effective theory
below the scale $\LS$ at which SUSY breaking is communicated \footnote{This scale should not be confused with the scale of SUSY breaking in the hidden sector. For instance, in the case of gravity mediation $\Lambda_{\rm hid} \simeq 10^{11}$ GeV while $\Lambda_{\rm SUSY} \simeq 10^{19}$ GeV.} to the
observable sector by an appropriate mediation mechanism. 
Likewise, in a theory addressing the origin of flavor, we can define a scale 
$\Lf$ characterizing the symmetry breaking scale at which the flavor structures arise. In the effective theory below $\Lf$, Yukawa couplings are generated as functions of the flavon vevs, $\langle\phi\rangle/\Lf$.

A relative hierarchy between $\LS$ and $\Lf$ can lead to flavor violating effects in the soft terms. For example, in the case of gravity mediation we expect $\Lf\lesssim M_{\text{Pl}} = \LS$. In this case the operators giving rise to the soft-breaking
terms are already present below $M_{\text{Pl}}$, while the flavor
symmetry is still exact at scales larger than $\Lf$. These operators must respect the family symmetries at the scale at which they are generated. However, at energies lower than
$\Lf$, in addition to the effective Yukawa couplings, corrections to these operators, and hence to the soft
breaking terms, are also generated as functions of $\langle \phi \rangle/\Lf$.
In the $\Lf \lesssim \LS$ case, we therefore expect new sources of 
flavor non-universality in the soft terms.\footnote{On the other hand, if $\LS\ll \Lf$, the soft terms are not present at
the scale of flavor breaking and SUSY breaking.  At $\Lf$ the flavor
interactions are integrated out but SUSY  still remains unbroken. The
only renormalizable remnant of the flavor physics below $\Lf$ are the
Yukawa couplings. At the scale $\LS$ soft-breaking terms can feel the flavor
breaking only through the Yukawa couplings or via non-renormalizable operators  proportional to $\LS/\Lf$ and therefore are
negligible. One typical example of this scenario will be the case of gauge mediated supersymmetry breaking (GMSB)\cite{Giudice:1998bp,Raidal:2008jk}.}

In this article we will show that flavor non-universal couplings  are
unavoidable in the soft-breaking terms when $\Lf \lesssim
\LS$. To illustrate our point, we assume that SUSY breaking is primarily mediated by supergravity, 
and that the Yukawa structures arise from a broken flavor symmetry through the FN mechanism. We consider two instructive choices for the flavor symmetry, an Abelian $U(1)_f$ and a non-Abelian $SU(3)_f$.   

We begin in Sec.~\ref{overview} by providing a brief review of the necessary background. Through simplified examples, this section also serves to illustrate the main results of this work, namely that these effective theories generically introduce new sources of flavor non-universality to the scalar sector of the Lagrangian (soft terms). As an application of these results, we consider two instructive choices for the flavor symmetry, an Abelian $U(1)_f$ and a non-Abelian $SU(3)_f$ in Sec.~\ref{examples}. As these SUSY models predict the order of magnitude of the flavor violation expected at low-energies, we show how phenomenological bounds from flavor observables can be used to provide information on the expected scale of the soft terms  in Sec.~\ref{phenomenology}. We conclude in Sec.~\ref{conclusions} with an overview of our results and prospects for further phenomenological studies.


\section{Supersymmetry and Flavor}
\label{overview}

The Minimal Supersymmetric Standard Model (MSSM) Lagrangian is determined by the particle content (gauge representations), the superpotential and the soft supersymmetry-breaking terms. The MSSM economically contains only the supersymmetric field content of the SM, with an additional Higgs doublet. As its superpotential (up to the $\mu$-term) is  fixed by the Yukawa sector of the SM, the soft-breaking terms encode most of  the new parameters of the Lagrangian, unfortunately introducing a huge number of unknowns to the theory (a generic MSSM has a total of 123 parameters, 124 including $\theta_{QCD}$ \cite{Haber:1997if}). As these parameters can potentially give rise to dangerous flavor violation, the MSSM is often said to suffer from a SUSY flavor problem. 

It is possible however that the mechanism responsible for generating the soft-terms preserves some or all of the structure of a flavor-symmetric ultraviolet completion to the MSSM. As we demonstrate, this is precisely the scenario considered here in this work, where an underlying spontaneously-broken flavor symmetry is {\em simultaneously} responsible for the fermion masses and mixing angles and the different flavor structures in the soft-breaking terms. In this case, the new flavor violating couplings in the SUSY Lagrangian provide a magnificent opportunity to measure new data on flavor, and may hopefully provide hints towards a fundamental theory of flavor. 
 
As remarked in the introduction, supersymmetry must be broken in a hidden sector and transmitted to the visible sector through mediators, either radiatively or through non-renormalizable interactions. A minimal option to couple the hidden and visible sectors is gravity, typically referred to as a ``gravity-mediation''. This is the scenario we consider in this paper. In such a mediation scheme, non-renormalizable interactions with the hidden sector suppressed by $M_{\rm Pl}$ modify the couplings of the superpotential, $W$, and the kinetic terms specified by the K\"ahler potential\footnote{To specify completely the supergravity Lagrangian we would need also the gauge kinetic terms, but they play no role in the following discussion.}, $K$. But more importantly, the same non-renormalizable interactions also give rise to the soft-terms after SUSY is broken. 

Since we are mainly interested in the visible sector, and given that the couplings between the two sectors are gravitationally small, we can, in practice, neglect most of these interactions and the energy transfer between the hidden and visible sector. However, some effects of the hidden sector may still be important at low energies, especially if they break a symmetry which remains exact in the visible sector. An example of this is SUSY breaking; after SUSY is broken in the hidden sector
with a non vanishing F-term, $\langle F \rangle \neq 0$, the gravitational interactions transmit this breaking with a strength $m_{\rm soft} \simeq \langle F \rangle/M_{\text{Pl}}$ to the visible sector. The remnants of these interactions in the visible sector are soft SUSY-breaking masses for the scalars and gauginos, as well as the scalar trilinear terms 

In the following, we assume for simplicity that there is a single F-term, $F_X$, that encodes the effects of SUSY-breaking in the hidden sector, and couples, through gravitational interactions, {\em universally} to all the visible-sector fields. It is important to remember that, aside from pure visible sector couplings, gravitational couplings between the visible and hidden sectors are always present in $K$ and $W$ below $M_{\text{Pl}}$. These couplings are renormalized by visible interactions through RGEs up to the mediation scale $M_{\text{Pl}}$. In the case of a single F-term defined above, all the soft terms in the ``full'' theory, above the flavor scale, $\Lf$, are universal. However, if a flavor symmetry is indeed present in the theory, its pattern of breaking and the details of the underlying theory will dictate new non-universal flavor structures in the low-energy efective theory below $\Lf$. 

To illustrate this point, let us consider an Abelian flavor symmetry, $U(1)_f$, that breaks spontaneously at a scale $\Lf < M_{\text{Pl}}$, generating the flavor structures. Before the breaking of the flavor symmetry, our superpotential and K\"ahler potential are invariant under both the SM (or GUT) and flavor symmetries. Suppose that the relevant superpotential terms allowed by the symmetries of the theory and which couple the flavons and mediators to the SM fields is given by,
\bea
W = g \left( \psi_3 \bar \chi_{-2} \phi+ \chi_2 \bar \chi_{-1} \phi + \chi_1 \bar \chi_0 \phi +  \chi_0 \bar \psi_{0} H + \psi_0 \bar \chi_{-1} \bar \phi\right) + M \left( \chi_0 \bar \chi_0 + \chi_1 \bar \chi_{-1} + \chi_2 \bar \chi_{-2} \right)  + m \phi \bar \phi + \ldots \,,
\label{mediatorW}
\eea
where we have used subscripts to denote the  $U(1)_f$-charges of the SM and mediator fields, and the flavon $\phi$~($\bar \phi$) is taken to have charge $-1$ ($+1$).  We assume a common coupling $g$ and mediator mass $M$ to simplify the discussion.

\begin{figure} 
\center
\includegraphics[width=0.7\textwidth]{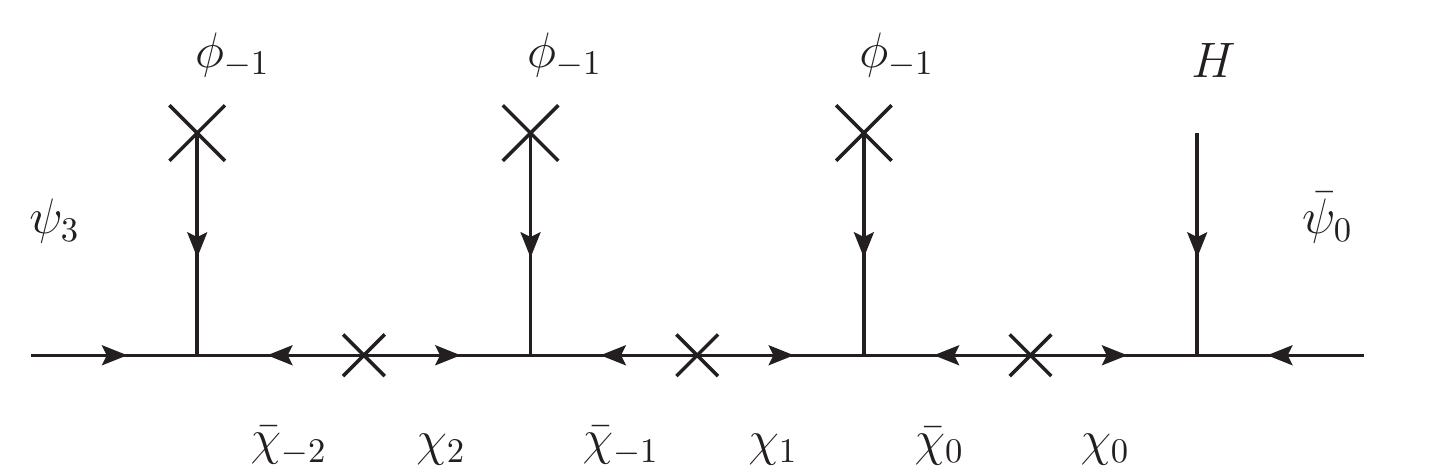}   %
\caption{Contribution to the superpotential in superfield notation. The superpotential is holomorphic and therefore involves only fields and not daggered fields, \emph{i.e.}, all arrows must enter the vertices in superfield notation.}
\label{fig:superpot1}
\end{figure}

Assuming that at $\Lambda_{ f} \sim \langle \phi \rangle$ the flavor symmetry is spontaneously broken, a Yukawa term will be generated upon integrating out the mediators, as shown in the supergraph of Fig.~\ref{fig:superpot1},\footnote{In superfield notation arrows pointing towards the vertex correspond to left-handed fields and those leaving from the vertex right-handed, or alternatively daggered, fields. Then the superpotential terms are vertices with all arrows entering  ($W$) or leaving the vertex ($W^*$) while K\"ahler vertices have arrows both entering and leaving the vertex\cite{Grisaru:1979wc,Gates:1983nr,Drees:2004jm}.} which gives rise to a contribution in the effective superpotential,
\bea
W_{\rm eff} = g^4 \left(\frac{\langle \phi \rangle}{M}\right)^3 \psi_3 \bar \psi_{0} H +~ \ldots \,.
\label{effW1}
\eea

\begin{figure} 
\center
\includegraphics[width=0.8\textwidth]{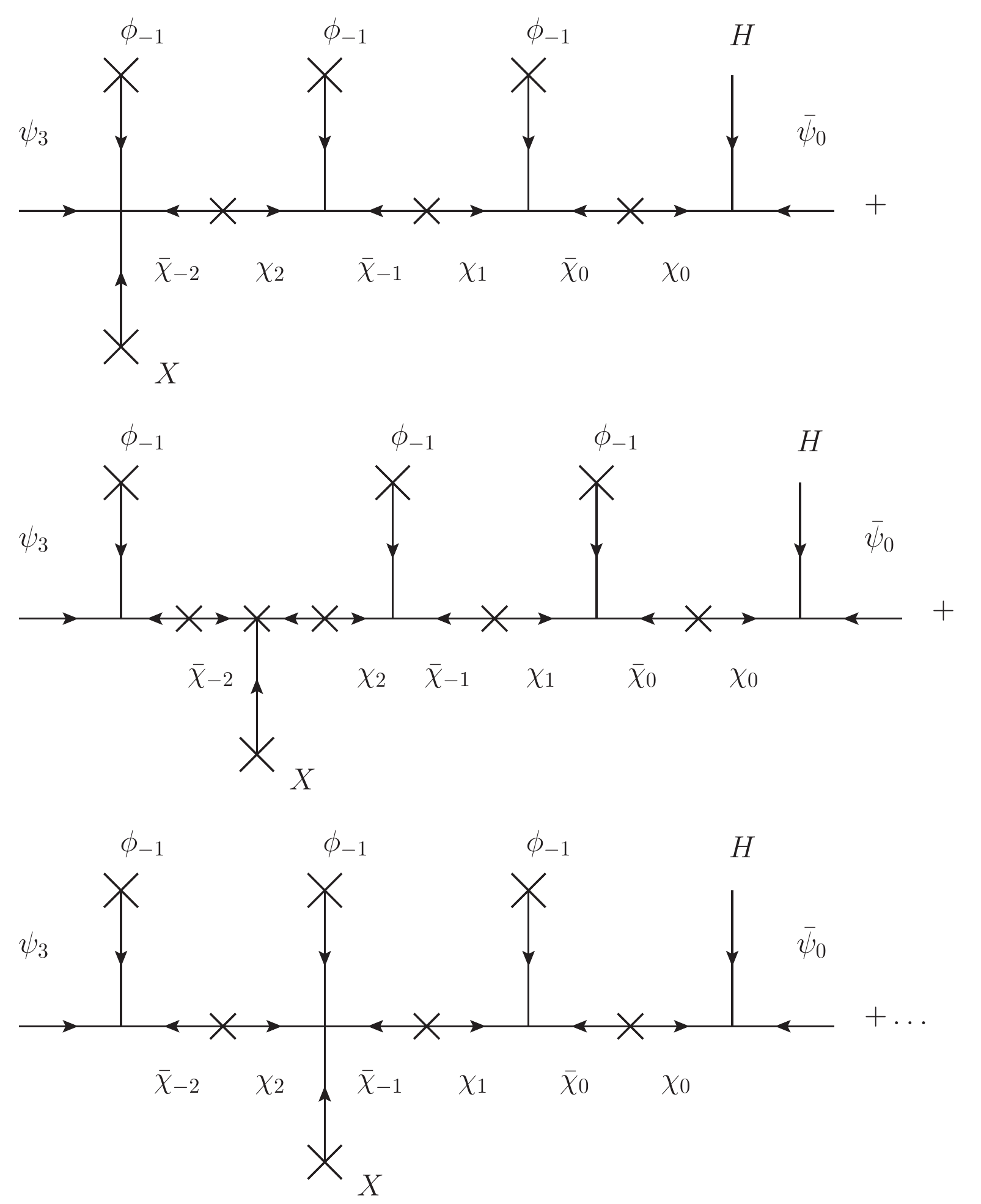}   %
\caption{Contributions to the A-terms from Fig.~\ref{fig:superpot1}. As noted, for this diagram seven insertions of $F_X$ are possible.}
\label{fig:superpot2}
\end{figure}

Further flavor structures will be generated in the soft-breaking terms, obtained from the superpotential in the full supergravity theory after SUSY is broken in the hidden sector.  Above the flavor breaking scale, universal soft terms are generated. For instance, trilinear terms in the potential of the ``full'' theory are simply proportional to the superpotential, \eq{mediatorW}, as V = $m_{3/2} \times W$. In terms of the spurion $X$, which encodes the breaking of SUSY, these trilinear terms correspond to the non-renormalizable couplings $\frac{X}{M_{\rm Pl}} \times W$ with an F-term replacing the $X$ field.

Similarly, new trilinear terms will be generated in the  effective potential $V_{\rm eff}$ below the flavor breaking scale. However, unlike the contributions to $W_{\rm eff}$, it is easy to see that the trilinear scalar coupling in the effective theory at low energies corresponding to the effective Yukawa coupling of Fig.~\ref{fig:superpot1} has seven different contributions; four with $F_X$ inserted in any of the cubic vertices, plus three contributions with $F_X$ in the superpotential mass, as shown in Fig.~\ref{fig:superpot2}. Here, we have used the fact that all vertices in the full superpotential have a corresponding trilinear coupling stemming from the non-renormalizable interaction $\frac{X}{M_{\rm Pl}} \times W$, and each gives a contribution to the effective trilinear term in the low energy theory of equal size.
\begin{figure}
\center
\includegraphics[width=\textwidth]{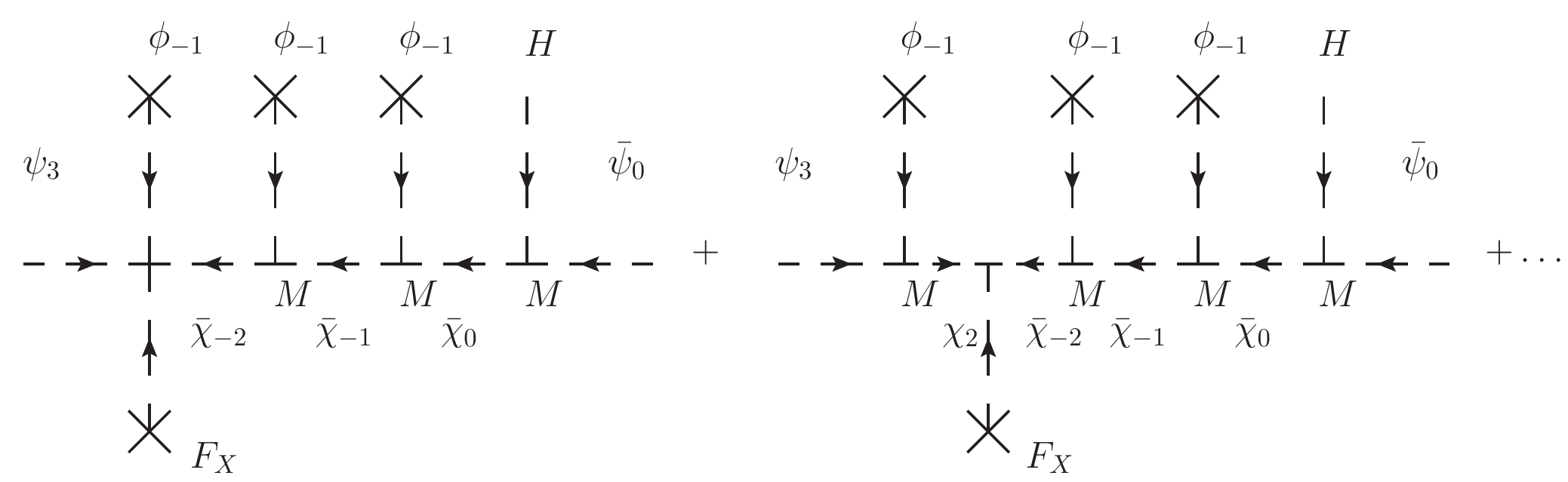}   %
\caption{Fig.~\ref{fig:superpot2} drawn in terms of the scalar component fields.}
\label{fig:aterm}
\end{figure}

The same diagrams drawn in terms of the component scalar fields are shown in Fig.~\ref{fig:aterm}. The vertices proportional to $M$ are obtained from the interference of the different components of the $F$-terms  from \Eqn{mediatorW}, which are 
 $|F_{\chi_{2}}|^2$, $|F_{\chi_{1}}|^2$ and $|F_{\chi_{0}}|^2$. The term $F_X/M_{\rm Pl}= m_{3/2}$ corresponds to the trilinear term $\frac{X}{M_{\rm Pl}} \times W$. 

Integrating out the heavy fields and replacing the scalar propagators by $1/M^2$, we obtain 7 identical contributions with coupling
$m_{3/2} (\langle \phi \rangle /M)^3$, generating a term $A_{3,0}= 7 \times m_{3/2} (\langle \phi \rangle/M)^3$ in the effective potential; this is to be compared with the Yukawa coupling of the effective superpotential, for which the single supergraph generates only $Y_{3,0}=
(\langle \phi \rangle /M)^3$. By this simple example, it is clear that the proportionality factor will
change depending on the number of vertices in the full theory and
therefore, if we have a different number of flavon insertions generating the various Yukawa elements, {\em the trilinear matrices will never be proportional to Yukawa matrices} \cite{Ross:2002mr,Calibbi:2008qt,Calibbi:2009ja}.

\begin{figure} 
\center
\includegraphics[width=0.3\textwidth]{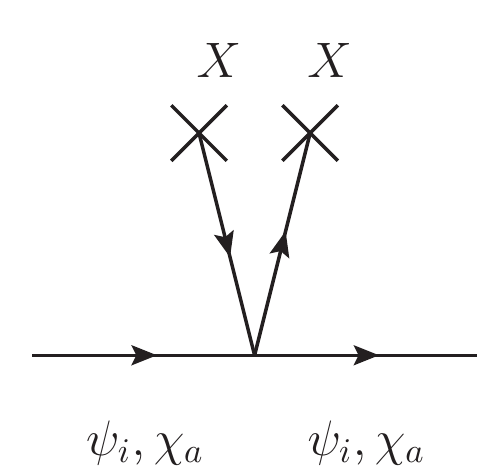}   %
\caption{Contribution to the soft mass from the K\"ahler potential in superfield notation. Arrows entering the vertex represent fields, while arrows leaving the vertex are daggered fields; here $|\langle F_X \rangle |^2/M_{\text{Pl}}^2 = m_{\rm 3/2}^2$.}
\label{fig:smass}
\end{figure}

A similar mismatch occurs for the soft masses. Before flavor symmetry breaking, the soft-masses, obtained from the K\"ahler potential, are universal for all visible fields. This includes the SM, flavon  and any mediator fields needed in the model. In terms of the spurion $X$, we can represent these flavor symmetric soft-terms through the supergraph of  Fig.~\ref{fig:smass}, and the K\"ahler function at this point would be,   
\bea
K = \psi^\dagger_i \psi_j \left( \delta_{i j} + \delta_{i j} \frac{ X X^\dagger}{M_{\rm Pl}^2} \right)\, ,
\eea  
with $\psi_i$ any of the visible fields.

\begin{figure}
\center
\includegraphics[width=0.5\textwidth]{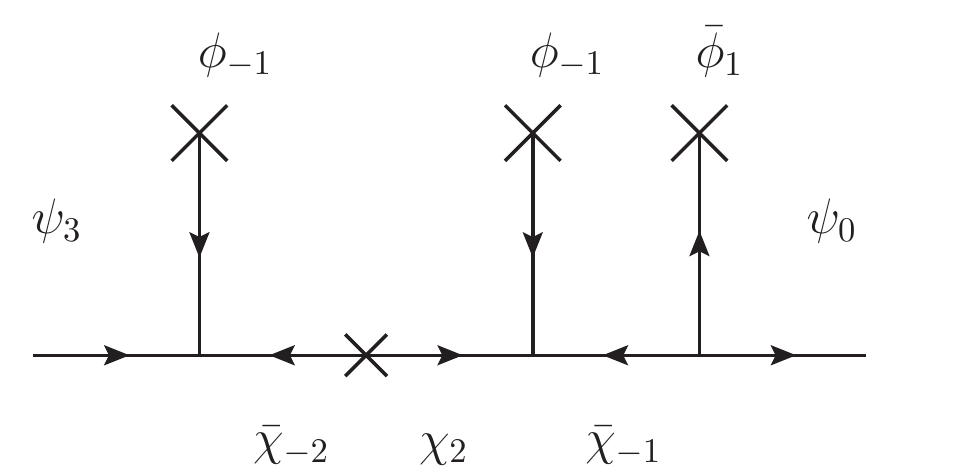}   %
\caption{Supergraph illustrating a possible non-diagonal K\"ahler couplings induced by integrating out the heavy mediator superfields.}
\label{fig:kahler1}
\end{figure}

 As was the case for the superpotential, after the flavor symmetry is broken, the K\"ahler potential receives new contributions. Since the K\"ahler potential includes the usual wave-function renormalization,  the
 light-field two-point functions in the low energy effective theory receive new contributions mediated by the heavy fields, coupling fields with different flavor charges through an appropriate number of flavon insertions.  An example of these two-point functions is shown in
 Fig.~\ref{fig:kahler1}. In this case, field redefinitions will be necessary to ensure
 canonical kinetic terms, see \cite{King:2004tx}.

\begin{figure}
\center
\includegraphics[width=\textwidth]{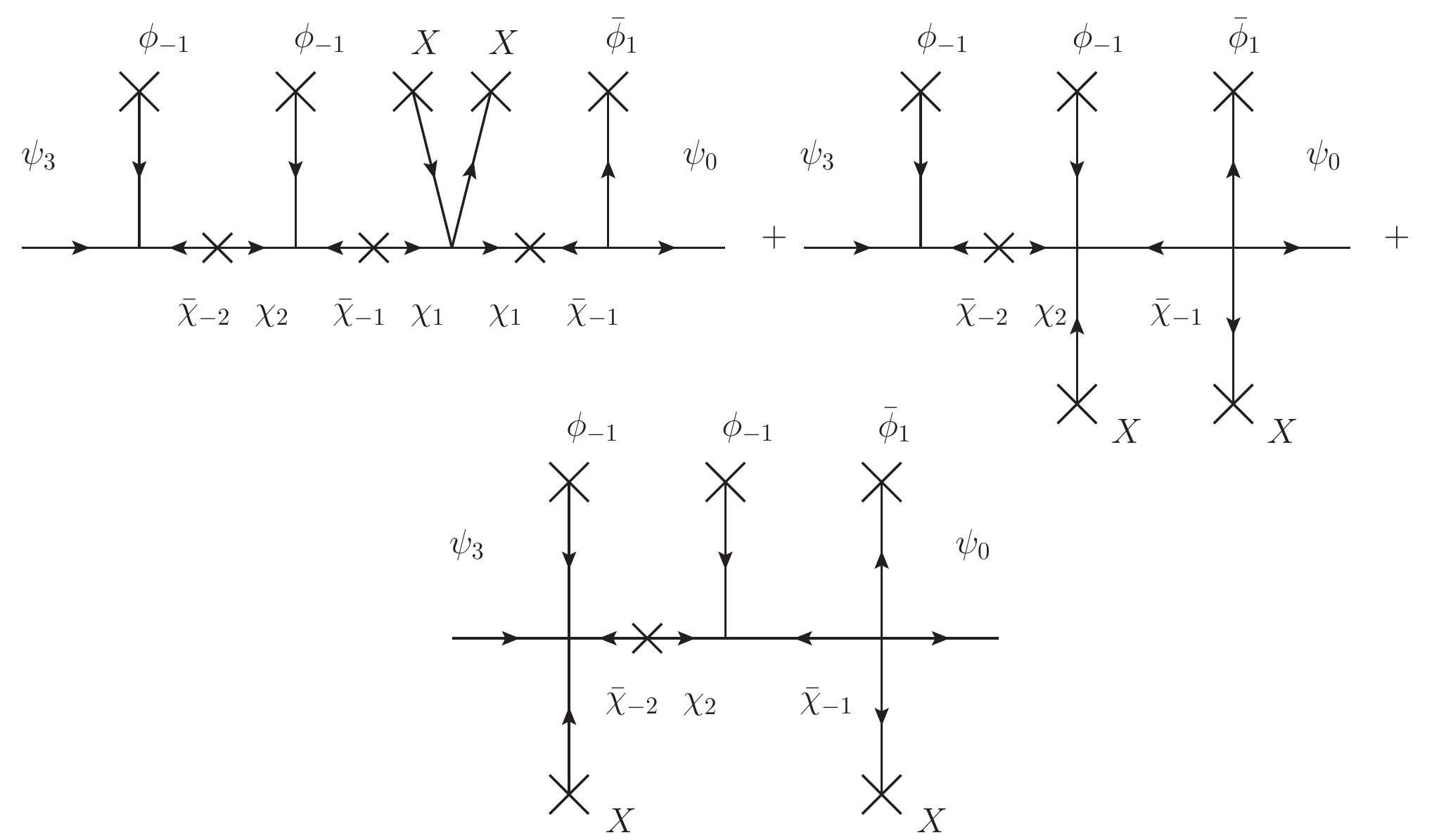}   %
\caption{Off-diagonal contributions to the scalar soft-masses from the K\"ahler diagram of Fig.~\ref{fig:kahler1}. Again, multiple possibilities exist for inserting the supersymmetry breaking term $X  X^\dagger $.}
\label{fig:kahlermass}
\end{figure}

Following the same logic as for the trilinear interactions, inserting the SUSY breaking F-term in all possible ways at different points in the diagram makes a difference between the kinetic terms and the 
soft masses. Again we have several diagrams contributing equally to the soft-breaking masses for each diagram 
renormalizing the kinetic terms, Fig.~\ref{fig:kahlermass}. In general, the K\"ahler potential after integrating the heavy mediator fields is now, 
\bea
\label{kahlereff}
K = \psi^\dagger_i \psi_j \left( \delta_{i j} + c~\left(\frac{\langle \phi \rangle}{M_\chi}\right)^{q_{ij}} + \left(\delta_{i j}+ b~\left(\frac{\langle \phi \rangle}{M_\chi}\right)^{q_{ij}}\right) \frac{ X X^\dagger}{M_{\rm Pl}^2} + {\rm h.c.} \right)
\eea 
with $b \sim N c$, where $N$ is the number of internal lines we can insert the spurion field.\footnote{As in the case of the trilinears, this may be expressed in terms of the component fields.} Given that the kinetic terms and the soft masses are not proportional, the field redefinition necessary to ensure canonical kinetic terms will not simultaneously diagonalize the soft masses, introducing new sources of flavor violation. 

These results, that trilinear interactions and soft masses are not proportional in the low-energy theory after integrating out heavy fields, apply generically to all supersymmetric effective theories of flavor with $\Lf \lesssim \LS$. 

\section{Flavor Symmetries}  \label{examples}

As a first application of the results of the previous section, we consider now the effects of integrating out heavy mediator fields on the structure of the soft-breaking terms in two representative flavor models: a toy $U(1)_f$ model, as an example of an Abelian symmetry, and a non-Abelian $SU(3)_f$ model.

In the following, we assume the conditions on the breaking of SUSY of the previous section apply: i) gravity mediation and ii) a single F-term with universal couplings is responsible for generating all the soft-terms of the visible sector. Our goal is not to present completely realistic models, but to show that non-universal soft-terms are a generic prediction of these models, even starting with fully universal soft terms at the mediation scale. Realistic models will have to deal with other problems like Golstone bosons in the case of global symmetries or D-flatness in the case of gauged non-Abelian symmetries, etc. Solutions to all these problems can be found in the literature\cite{Ross:2004qn,Babu:2014sga} or can be avoided altogether with discrete flavor symmetries\cite{Ishimori:2010au,Altarelli:2010gt,King:2013eh}.  

\subsection{Toy $U(1)_f$ Model} \label{sec:toy}

Under the conditions outlined above, our toy $U(1)_f$ model is completely defined by its superpotential, which generalizes the superpotential of Eq.~(\ref{mediatorW}),
\bea 
W &\supset& g \sum_{q} \left(  \psi_{q}^{} \bar \chi_{-q +1}^{}\phi + \chi_{q} \bar \chi_{-q+1}^{} \phi + \chi_{q -1}^{} \bar \chi_{-q}^{} \bar \phi + \chi_{-q}^{} \bar \psi_{q}^{u} H_u + \chi_{-q}^{} \bar \psi_{q}^{d} H_d \right) + 
\nonumber \\
&& +g\left( \bar \psi_{q}^{u} \bar \chi_{-q +1}^{u}\phi  + \chi_{q}^{u} \bar \chi_{-q+1}^{u} \phi + \chi_{q -1}^{u} \bar \chi_{-q}^{u} \bar \phi  + \bar \psi_{q}^{d} \bar \chi_{-q +1}^{d}\phi  + \chi_{q}^{d} \bar \chi_{-q+1}^{d} \phi + \chi_{q -1}^{d} \bar \chi_{-q}^{d} \bar \phi  \right) + \nonumber \\
&& M  \sum_{q}  \left( \chi_{q}^{} \bar \chi_{-q}^{} +  \chi_{q}^{u} \bar \chi_{-q}^{u} + \chi_{q}^{d} \bar \chi_{-q}^{d} \right)+ m \phi \bar \phi \,,
\eea
where the $\psi$s ($\bar \psi$s) represent the left-handed (right-handed) SM superfields, $\chi$s and $\bar \chi$s the mediators, and $\phi$ and $\bar \phi$ are the flavons. As before, the subscripts for the $\psi$s and $\chi$s denote their respective $U(1)_f$ charges, chosen positive for the SM fields. The flavon, $\phi$ ($\bar \phi$) is assigned to have $U(1)_f$ charge $-1$ ($+1$). For simplicity, we assume a common coupling, $g$, and a common mass, $M$, for all the mediators. In principle, however, each mediator particle can couple with its own $\order(1)$ coupling and can have a different mass. Similarly, it is well-known that one of the main problems of supersymmetric $U(1)_f$ models is that the equality of soft-masses of the different generations, even in the symmetric limit, is not guaranteed by symmetry. This inequality may generate  large off-diagonal entries after rotating to the basis of diagonal Yukawa matrices. This problem is absent in our toy model due to our assumption of a single $F$-term with universal couplings to the visible sector, but should be addressed in a realistic Abelian model.

To obtain the low-energy Yukawa couplings and the soft-breaking trilinear interactions in the effective theory below $\Lf$ (or the soft-masses and the K\"ahler function), it is again easiest to work with supergraphs. Drawing supergraphs such as that of Fig.~\ref{fig:superpot1}, one finds that in terms of the small expansion parameter, $\varepsilon ~=~ g (\langle \phi \rangle / M ) ~\ll~ 1$,
integrating out the heavy messengers will generate Yukawa interactions of the form
\beq 
 Y_{ij}^{} ~\sim ~ \varepsilon^{n_{ij}}_{}, \qquad n_{ij}^{} = q_i + q_j \,.
 \eeq
 
For the trilinear couplings, we have to take into account that each supergraph generating the Yukawa terms will necessarily contain $n_{ij}$ flavon insertions (cubic vertices) and $n_{ij}$ messenger masses, with one additional vertex to couple the Higgs. This will result in $2 n_{ij}+1$ possibilities to insert the SUSY breaking $F_X$ term, so that the trilinear interactions are given by,
\beq 
A_{ij}^{} ~\sim ~ m_{3/2}^{}~(2 n_{ij} +1)~ Y_{ij}^{}.
\eeq 
The distinct prefactors, $\sim n_{ij}$, in the entries of the trilinear matrix make it {\em not proportional} to the Yukawa matrix, as advertised. 

We can apply this to an explicit $SU(5)$-inspired pattern for the $U(1)_f$ charges of the SM fermions: $(Q_3, 0)$, $(Q_2, 2)$,  $(Q_1, 3)$,  $(\overline{d}_3, 0)$, $(\overline{d}_2, 0)$, $(\overline{d}_1, 1)$, $(\overline{u}_3, 0)$, $(\overline{u}_2, 2)$, $(\overline{u}_1, 3)$: Given these charges, the Yukawa matrices are given by,  
\bea
\label{theYmatrixI}
Y_{u} \sim y_t \left(
\begin{array}{ccc}
e \varepsilon^6&\varepsilon^5&  \varepsilon^3\\
\varepsilon^5&d \varepsilon^4&  \varepsilon^2\\
\varepsilon^3& \varepsilon^2& 1
\end{array}
\right), \qquad\qquad  Y_{d} \sim y_b \left(
\begin{array}{ccc}
e'\varepsilon^4&f'\varepsilon^3& f\varepsilon^3\\
\varepsilon^3&d'\varepsilon^2& h \varepsilon^2\\
\varepsilon &k'& k
\end{array}
\right),
\eea
where we have taken the liberty to add $\order(1)$ coefficients in the different entries to reproduce the observed masses and mixings.\footnote{This freedom is a general feature of Abelian models, and can be motivated by modifying our simplified superpotential to allow distinct $\order(1)$ couplings for each term.} Using $\varepsilon = \lambda_C = 0.225$ as the expansion parameter and choosing $e= 1.1$, $e'=1.8$, $d=4.0$, $d'=0.90$, $f=1.3$, $f'=0.82$, $h=1.3$, $k=0.69$ and $k'=1.2$ gives a good fit to the measured quark masses, reproduces correctly the Cabibbo block of the CKM matrix and approximates the smaller CKM elements.

Assuming the same $\order(1)$ coefficients appear in the graphs generating the trilinear couplings, we would obtain 

\bea
\label{theAmatrixI}
 A_{u} \sim m_{3/2}~y_t \left(
\begin{array}{ccc}
13 e\varepsilon^6& 11 \varepsilon^5& 7 \varepsilon^3\\
11 \varepsilon^5& 9 d \varepsilon^4& 5 \varepsilon^2\\
7  \varepsilon^3& 5 \varepsilon^2& 1
\end{array}
\right), \qquad\qquad  A_{d} \sim m_{3/2}~y_b \left(
\begin{array}{ccc}
9 e'\varepsilon^4& 7 f'\varepsilon^3& 7f \varepsilon^3\\
7 \varepsilon^3& 5 d' \varepsilon^2& 5 h \varepsilon^2\\
3  \varepsilon &k'& k
\end{array}
\right).
\eea
To get a sense for the flavor violation induced by this mismatch in order one coefficients, we must go to the Super-CKM (SCKM) basis\footnote{In principle, we should first redefine the fields and obtain canonical kinetic terms as discussed below. However, it can be shown that these fiels redefinitions introduce only subleading corrections in $\varepsilon$\cite{King:2004tx}.}, where the Yukawas are diagonal and
in which $A_u$ and $A_d$ take the form,

\begin{equation}
\label{trilinU1}
  \nonumber  
A_u^{} \rightarrow   U_u^{} A_u V_u^{\dagger} ~ \approx ~ 
m_{3/2}~y_t \left(
\begin{array}{ccc}
 14.6 \varepsilon^6 & 10.9 \varepsilon^5 & 7 \varepsilon^3 \\
 10.9 \varepsilon^5 & 35.9 \varepsilon^4 & 4.9 \varepsilon^2 \\
 7 \varepsilon^3 & 4.9 \varepsilon^2 & 1 \\
\end{array}
\right),
\quad 
A_d^{} \rightarrow   U_d^{} A_d V_d^{\dagger} ~ \approx ~
m_{3/2}~y_b \left(
\begin{array}{ccc}
 11.3 \varepsilon^4 & 4 \varepsilon^3 & 6.5 \varepsilon^3 \\
 4.9 \varepsilon^3 & 3.2 \varepsilon^2 & 4.3 \varepsilon^2 \\
 2.1 \varepsilon & 0.86 & 0.49 \\
\end{array}
\right).
\end{equation}
We see that indeed large flavor violating entries can be expected in the trilinear matrices of the low energy effective theory, even starting with completely universal soft terms in the full theory. This non-universality can constrain strongly the allowed squark and gluino masses of the model. 

In addition to generating the Yukawa and trilinear interactions, integrating over the heavy mediators will also give new off-diagonal contributions to the  K\"ahler function,
\beq
\label{kahlerab}
K ~=~ \psi_i^{\dagger} \psi_j^{} \left( \delta_{ij}^{} + c_{ij}^{} ~+~ (\delta_{ij}^{} + b_{ij}^{})\frac{X X^\dagger}{M_{\rm Pl}^2} + \rm h.c. + \ldots \right),
\eeq 
where the coefficients $c_{ij}$ encode the supersymmetric off-diagonal entries in the kinetic terms while the $b_{ij}$ give rise to the soft-breaking masses; the ellipses denote terms of higher order. 

The mismatch between the wave-function renormalization ($\delta_{ij} + c_{ij}$) and the soft masses ($\delta_{ij} + b_{ij}$) can again be calculated by drawing the appropriate supergraphs. Looking at Fig.~\ref{fig:kahler1}, for example, and remembering that the super-propagator of the mediator gives a factor of $M^{-2}$ (left-right, uncrossed propagator) or $M^{-1}$ (left-left or right-right, crossed propagator), it is tempting to guess that for the K\"ahler function, the leading corrections are proportional to $\varepsilon^{q_{ij}}$, where $q_{ij}=|q_i -q_j|$. 

\begin{figure}
\center
\includegraphics[width=0.4\textwidth]{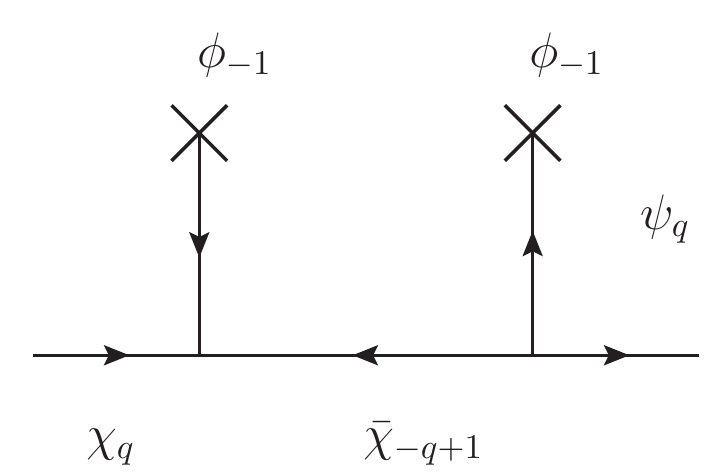}   %
\caption{Supergraph giving an effective coupling between the messengers and the SM fields. Such a term is always present but in general gives only higher order $\varepsilon$ corrections to the K\"ahler function.}
\label{fig:mixing2}
\end{figure}

This guess is indeed true, except for the special case of $q_{ij}=1$. Given that our superpotential does not include direct couplings between the SM superfields $\psi_{q}$ and the mediators $\bar \chi_{-q}$,  one can only generate such a mixing through the supergraph of Fig.~\ref{fig:mixing2}, at the cost of a higher $\varepsilon$ dependence than naively expected. 
Taking this subtlety into account, we find explicitly that 
\begin{equation}
\label{eq:wavefunc}
c_{ij}^{} \sim \begin{cases}
                               (q_{ij}^{} - 1)\varepsilon^{q_{ij}} &\text{for $ q_{ij}^{} \geq 2$} \\
                               2 \varepsilon^3 &\text{for $q_{ij}^{} = 1$} \\
                               \varepsilon^2 &\text{for $q_{ij}^{} = 0$} \,.
                       \end{cases}
\end{equation}
where we included the leading correction to the diagonal terms that comes from the diagram with a single mediator super-propagator and two $\phi$'s, while the form of the coefficient for the off-diagonal terms stems from the number of different diagrams that can be drawn which contain only a single left-right super-propagator for the mediators.

For the soft masses, as remarked in the previous section, we will have $b_{ij} = N c_{ij}$, where $N$ is the number of SUSY breaking insertions, $X^\dagger X/M_{\rm Pl}^2$, we can make for a given supergraph. The counting for the number of possible insertions is straightforward.
For a given supergraph contributing to the K\"ahler function, without increasing the powers of the flavor mediator mass, $M$, in the denominator, we can only insert $X^\dagger X/M_{\rm Pl}^2$ in the left-right propagator. However, there is also the possibility of inserting $X^\dagger$ in the single $W^*$ vertex, with $(q_{ij} -1)$ options to insert the $X$ in the remaining $W$ vertices (see Fig.~\ref{fig:kahlermass}). This gives $N = (q_{ij}-1) +1 = q_{ij}$ possibilities. Again, the subtlety for the case $q_{ij} = 1$ requires the additional mass insertion of Fig.~\ref{fig:mixing2}, so that in total we find,
\begin{equation}
\label{eq:smass}
N = \begin{cases}
                 q_{ij}^{} & \text{for $ q_{ij}^{} \geq 2$} \\
                 3 & \text{for $q_{ij}^{} = 1$} \\
                 2 & \text{for $q_{ij}^{} = 0$} \,.
       \end{cases}
\end{equation}

An explicit example, using the same $SU(5)$-inspired charge assignment as before, would give the same coefficients $c_{ij}$ and $b_{ij}$ for $Q$, $\overline{u}$ and $\overline{d}$, 
\beq
c_{ij}^{(Q,\overline{u},\overline{d})} ~=~ \left( \begin{array}{ccc}
\varepsilon^2 & 2 \varepsilon^3 & 2 \varepsilon^3 \\
2 \varepsilon^3 & \varepsilon^2 & \varepsilon^2   \\
2 \varepsilon^3 & \varepsilon^2 & \varepsilon^2 
\end{array} \right),
 \qquad\qquad
b_{ij}^{(Q,\overline{u},\overline{d})} ~=~ \left( \begin{array}{ccc}
2 \varepsilon^2 & 6 \varepsilon^3 & 6 \varepsilon^3 \\
6 \varepsilon^3 & 2 \varepsilon^2 & 2 \varepsilon^2   \\
6 \varepsilon^3 & 2 \varepsilon^2 & 2 \varepsilon^2 
\end{array} \right).
\eeq

Again, they are not proportional. One can see this explicitly by canonically normalizing the fields and diagonalizing the matrices $\delta_{ij} + c_{ij}$. As remarked in \cite{King:2004tx}, this can always be achieved by an upper-triangular matrix $T$, $\psi^{\prime} = T \psi$, 

\beq
T ~=~ \left( \begin{array}{ccc}
1+\frac{\varepsilon^2}{2} & 2 \varepsilon^3 & 2 \varepsilon^3 \\
0 & 1+\frac{\varepsilon^2}{2} & \varepsilon^2   \\
0 & 0 & 1+\frac{\varepsilon^2}{2}
\end{array} \right).
\eeq 
Aside from giving higher-order $\varepsilon$ corrections to the Yukawa matrices, $Y \rightarrow (T^{-1})^\dagger Y T^{-1}$, this will give the soft masses in the canonical basis, $(m_{\rm soft}^2)_{ij} \equiv m_{3/2}^2 (\delta_{ij}+b_{ij})$,

\beq
\label{msoftU1}
(m_{\rm soft}^2)_{ij} \rightarrow (T^{-1})^\dagger (m_{\rm soft}^2)_{ij} T^{-1} \sim 
m_{3/2}^2 \left( \begin{array}{ccc}
1 +\varepsilon^2 - \frac{7}{4} \varepsilon^4 & 4 \varepsilon^3 & 4 \varepsilon^3 \\
4 \varepsilon^3 & 1 +\varepsilon^2 - \frac{7}{4} \varepsilon^4 & \varepsilon^2 -\frac{7}{4} \varepsilon^4   \\
4 \varepsilon^3 &  \varepsilon^2 -\frac{7}{4} \varepsilon^4 & 1 +\varepsilon^2 - \frac{19}{4} \varepsilon^4
\end{array} \right),
\eeq 
and due to the mismatch in the order one coefficients of $c_{ij}$ and $b_{ij}$, off-diagonal entries in the soft masses remain after canonical normalization.\footnote{To compare with the usual mass insertion bounds present in the literature, one would need to go to the SCKM basis, \emph{i.e.}, the basis of diagonal Yukawa couplings.}


\subsection{$SU(3)_f$ Model}

The situation is slightly different in the case of non-Abelian symmetries such as $SU(3)_{\it f}$, with flavon fields transforming as either a ${\bf 3}$ or ${\bf \bar 3}$. In this case, we must introduce several species of mediator fields, both singlets and triplets under $SU(3)_{\it f}$. We consider only $SU(2)_L$ singlet mediators, which always couple through a flavon to the right-handed SM fields and through the Higgs to the left-handed fields. 

As an illustrative example, we take the model of I.~de Medeiros Varzielas and G.~G.~Ross in Ref.~\cite{deMedeirosVarzielas:2005ax}, supplemented by the appropriate mediator sector. With the fields specified in Tables \ref{tab:higgsb} and \ref{tab:higgsb2} from \cite{deMedeirosVarzielas:2005ax}, the superpotential would be
\bea
W &=& g \left( \psi_i~ \bar\chi_{4}^i~ H + ~ \chi_{-4,i}~ \bar\chi_1 ~\bar \phi_{123}^i + ~ \chi_{-4,i}~\bar \chi_2 ~\bar \phi_3^i + ~\chi_{-4,i}~\bar \chi_3 ~\bar \phi_{23}^i  + ~\chi_{-3}~ \bar \chi_1~ \Sigma~ \right. \nn \\
& & ~~~ \left.  + ~ \chi_{-3} ~\bar \psi_i~ \bar \phi_{123}^{i}~ + ~ \chi_{-1} ~\bar \psi_i~ \bar \phi_{23}^{i}~ + ~ \chi_{-2} ~\bar \psi_i~ \bar \phi_3^{i}~+ ~\bar\chi_{3}~ \chi_{-5,i} ~ \bar \phi_{3}^{i}~  \right. \nn \\ 
& & ~~~ \left. +  ~\bar \chi_{2}~\chi_{-5,i}~ \bar \phi_{123}^{i}~  + ~\bar\chi_{1} ~\chi_{-3,i}~ \bar \phi_{3}^{i}~ + ~\bar \chi_{2}~\chi_{-3,i} ~ \bar \phi_{23}^{i}~\right) \nn \\
& & +~ M_u \left( ~\chi^u_1 ~\bar \chi^u_{-1} + ~ \chi^u_2 ~\bar \chi^u_{-2} + ~\chi^u_3 ~\bar \chi^u_{-3}~ + ~\bar \chi_3^{u,i} ~\chi^u_{-3,i}~+ ~\bar \chi_4^{u,i} ~\chi^u_{-4,i}~ + ~\bar \chi_5^{u,i} ~ \chi^u_{-5,i} \right) \nn \\
& & +~ M_d \left( ~\chi^d_1 ~\bar \chi^d_{-1} + ~ \chi^d_2 ~\bar \chi^d_{-2} + ~\chi^d_3 ~\bar \chi^d_{-3}~ + ~\bar \chi_3^{d,i} ~\chi^d_{-3,i}~+ ~\bar \chi_4^{d,i} ~\chi^d_{-4,i}~ + ~\bar \chi_5^{d,i} ~ \chi^d_{-5,i} \right) \,,
\eea

\begin{table}[tbp] 
\begin{tabular*}{1.00\textwidth}{@{\extracolsep{\fill}}|c|c c c c c c c|}
\hline
~${\bf Field}$ & $\psi$ & $\bar{\psi}$ & ~~~$H$ & $\Sigma$ &
$\bar\phi _{3}$ & $\bar\phi _{23}$ & $\bar\phi_{123}$~~~\\ 
\hline
~${\bf R }$ & $1$ &  $1$ & ~~~$0$ &  $0$ & $0$ & $0$ & $0$~~~\\
~${\bf U(1)}$ & $0$ & $0$ & $-4$ & $2$ & $2$ & $1$ & $3$~~~\\
\hline
~${\bf SU(3)_{\it f}}$ & ${\bf 3}$ & ${\bf 3}$ & ~~~${\bf 1}$ & ${\bf 1}$ & ${\bf \bar{3}}$ &
${\bf \bar{3}}$ & ${\bf \bar{3}}$~~~\\
\hline
\end{tabular*}
\caption{Transformation of the matter superfields under the $SU(3)$ family
symmetries.}
\label{tab:higgsb}
\end{table}

\begin{table}[tbp] 
\begin{tabular*}{1.00\textwidth}{@{\extracolsep{\fill}}|c|c c c c c c c c c c|}
\hline
~${\bf Field}$ & $\bar \chi_1$ & ~~$\chi_{-1}$ & ~~$\bar \chi_2$ & ~~$\chi_{-2}$ &~~ $\bar \chi_3$ & ~~$\chi_{-3}$ & ~~$\bar \chi_{4}^k$ & ~~$\chi_{-4,k}$ &~~ $\chi_{-3,k}$ &~~$\chi_{-5,k}$ ~\\ 
\hline 
~${\bf R }$ & $1$ & ~~~$1$ & $1$ & ~~~$1$ & $1$ & ~~~$1$ & $1$ & ~~~$1$ & $1$ &$1$\\
~${\bf U(1)}$ & $1$ & $-1$ & $2$ & $-2$ & $3$ & $-3$ & $4$ & $-4$ & $-3$ &$-5$~~~\\
\hline
~${\bf SU(3)}_{\it f}$ & ${\bf 1}$ & ~~~${\bf1}$ & ${\bf 1}$ & ~~~$\bf{1}$ & ${\bf 1}$ & ~~~${\bf 1}$ & ${\bf\bar 3}$ & ~~~${\bf  3}$ & ${\bf 3}$& ${\bf 3}$~~~\\
\hline
\end{tabular*}
\caption{Transformation of the mediator superfields under the $SU(3)$ family
symmetries.}
\label{tab:higgsb2}
\end{table}
suppressing the $u$ and $d$ indices on $\psi$ and $H$, and where the flavons, $\bar \phi_3$, $\bar \phi_{23}$ and $\bar \phi_{123}$, are labelled with a subscript indicating the field component whose vev is non-zero. The messenger subscripts denote their charges under an additional $U(1)$ present in the superpotential, and we neglect contributions from the $SU(2)_L$ doublet-mediators, which are assumed to be much heavier. 

The flavor symmetry is spontaneously broken in two steps: first $SU(3)_f\rightarrow SU(2)_f$, followed by the breaking of the residual $SU(2)_f$. We assume the following alignment of the flavon vevs,

\beq \nonumber
\langle \bar{\phi}_3 \rangle ~=~ (~0 ~~~ 0 ~~~ 1~) \otimes \begin{pmatrix}
~\alpha_u & 0 \\
0 & \alpha_d~ \end{pmatrix}, 
\eeq
\beq \nonumber 
\langle \bar{\phi}_{23} \rangle ~=~ (~0 ~~~ \beta ~ -\beta~),
\eeq
\beq
\langle \bar{\phi}_{123} \rangle ~=~ (~\gamma ~~~ \gamma ~~~ \gamma~~),
\eeq

where $\bar{\phi}_3$ transforms as a ${\bf 3} \oplus {\bf 1}$ under $SU(2)_R$, obtaining different vevs in the up ($\alpha_u$) and down ($\alpha_d$) sectors, while  $\bar{\phi}_{23}$ and $\bar{\phi}_{123}$ are $SU(2)_R$ singlets; the vevs are assumed hierarchical,  $\gamma \ll \beta \ll \alpha_u, \, \alpha_d$. 

\begin{figure}[h!]
\center
\includegraphics[width=\textwidth]{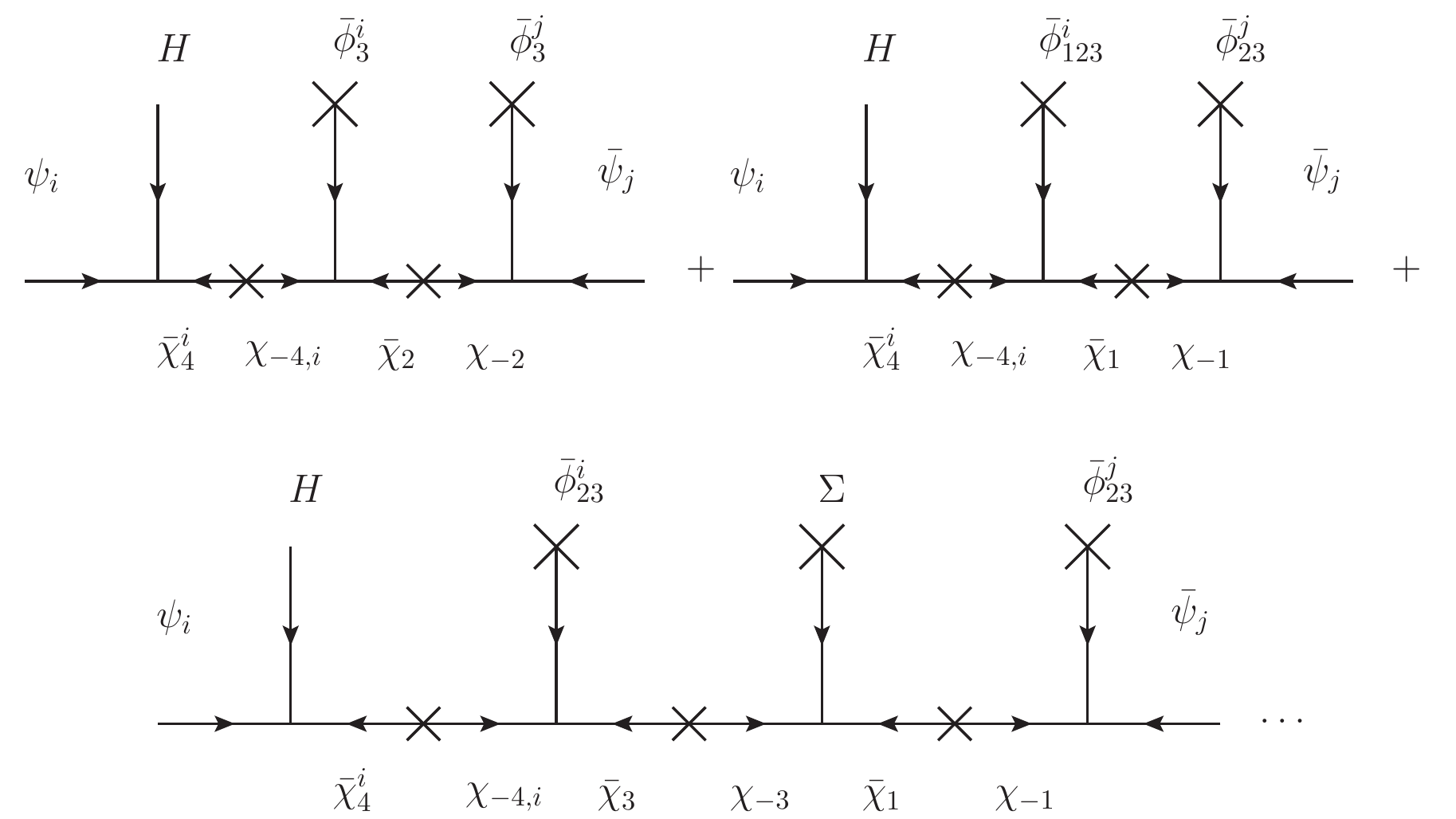}	\\
\caption{Contributions to the effective Yukawa couplings for the $SU(3)_f$ model in superfield notation.}
\label{fig:fig01}
\end{figure}

When the flavor symmetry is broken, the effective Yukawa couplings arise as the result of integrating out the heavy messengers in processes like those represented in Fig.~\ref{fig:fig01}. For instance, the dominant contribution to $Y_{33}$ will come from the first diagram, whereas the remaining terms in the $2\times3$ block will arise mostly from the third diagram. Similarly, $Y_{12}$ and $Y_{21}$ will be effectively generated by the second diagram and the analogous diagram obtained by changing the order of $\bar \phi_{23}$ and $\bar \phi_{123}$. The resulting Yukawa matrices are then

\beq
Y_u ~ \sim ~ y'_t 
\left(
\begin{array}{rrc}
0 & f~\varepsilon^2_u\,\varepsilon_d & -f^\prime~\varepsilon^2_u\,\varepsilon_d \\
f~\varepsilon^2_u\,\varepsilon_d & -\frac{2}{3}\,d\,\varepsilon_u^2 & ~~~\frac{2}{3}\,e\,\varepsilon_u^2 \\
-f^\prime~\varepsilon^2_u\,\varepsilon_d & \frac{2}{3}\,e\,\varepsilon_u^2 & 1
\end{array} 
\right),
\hspace{0.8cm}
Y_d ~ \sim ~ y'_b 
\left(
\begin{array}{rrc}
0 & h~\varepsilon^3_d & -k~\varepsilon^3_d \\h~\varepsilon^3_d & ~~l~\varepsilon_d^2 & -k^\prime~\varepsilon_d^2 \\
-k~\varepsilon^3_d & -k^\prime~\varepsilon_d^2 & 1
\end{array} 
\right),
\eeq

\begin{figure}[h!]
\center
\includegraphics[width=\textwidth]{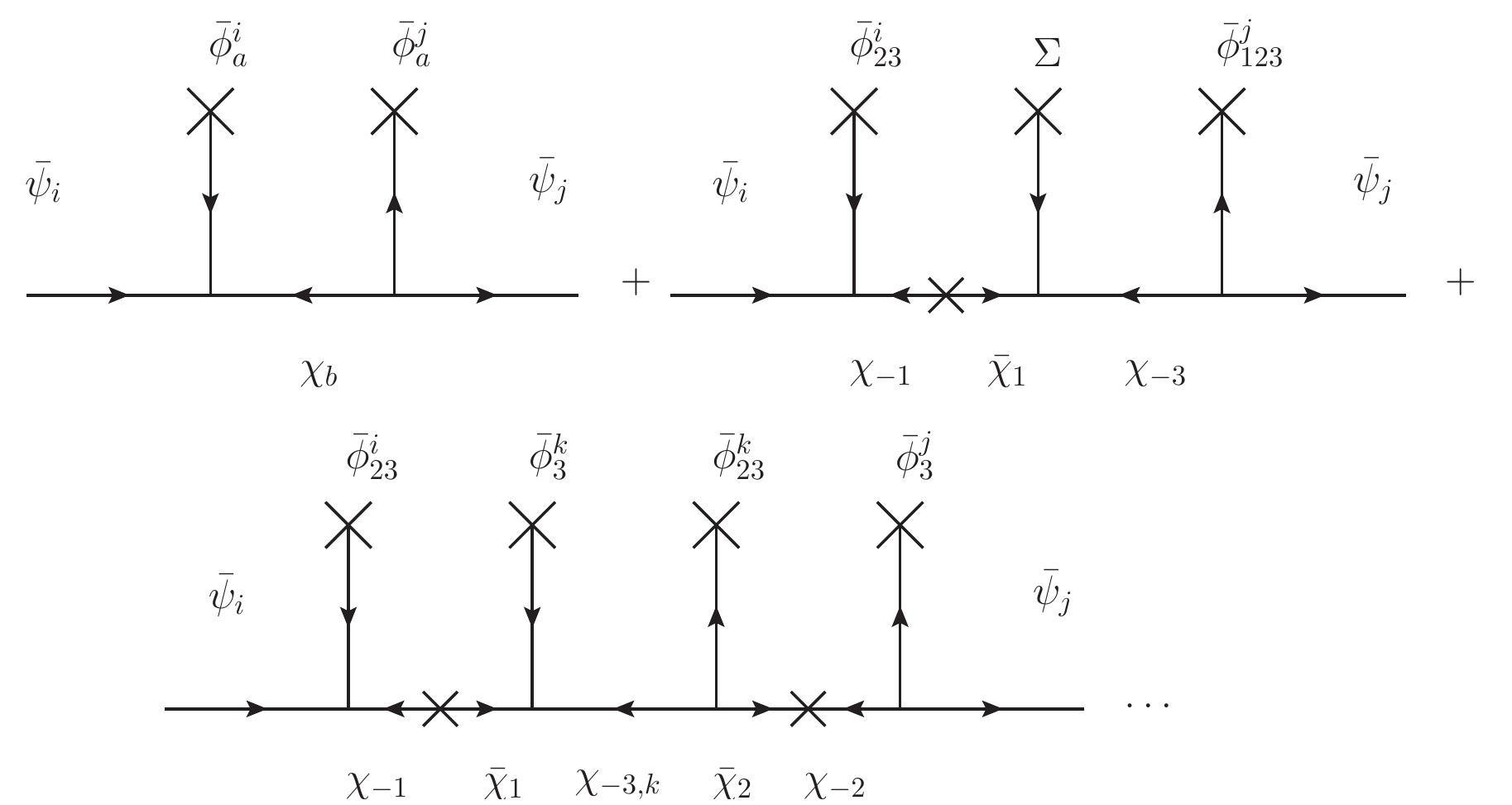} 
\caption{Contributions to the K\"ahler couplings in the $SU(3)_f$ model induced by integrating out the heavy mediator superfields.}
\label{fig:fig03}
\end{figure}

where $\frac{\alpha_u}{M_{u}} \simeq \frac{\alpha_d}{M_{d}} \simeq 0.7\equiv \alpha$, $\varepsilon_{f} \equiv \frac{\beta}{M_{f}}$, with $\varepsilon_{d}\simeq 3\,\varepsilon_{u} \sim 0.15$, $\frac{\gamma}{M_{d}} = \varepsilon_{d}^2 $ and $\langle\Sigma\rangle/M_{d}\simeq \mathcal{O}(1)$ whereas $\langle\Sigma\rangle/M_{u}\simeq -2/3\,\langle\Sigma\rangle/M_{d}$. The couplings $y'_t$ and $y'_b$ refer to the top and bottom Yukawa couplings before rotating the matrices to the canonical basis. Again, we allow for order one coefficients in the Yukawa matrices in order to reproduce quark masses and the CKM matrix. Here, they are chosen to be: $f\,=\,0.33$, $f^\prime \,=\,1.0$, $d\,=\,1.62$, $e\,=\,1.0$, $h\,=\,1.14$, $k\,=\,1.0$, $k^\prime\,=\,1.0$ and $l\,=\,0.74$. 

Once we know the diagrams which contribute to each effective Yukawa coupling, we may calculate the trilinear matrices, which are found to be
\beq
A_u \sim m_{3/2}~y'_t 
\left(
\begin{array}{crc}
0 & 5\,f~\varepsilon^2_u\,\varepsilon_d & -5\,f^\prime~\varepsilon^2_u\,\varepsilon_d \\
~~5\,f~\varepsilon^2_u\,\varepsilon_d & -\frac{14}{3}\,d\,\varepsilon_u^2 & ~\frac{14}{3}\,e\,\varepsilon_u^2 \\
-5\,f^\prime~\varepsilon^2_u\,\varepsilon_d & \frac{14}{3}\,e\,\varepsilon_u^2 & 5
\end{array} 
\right)
\hspace{0.2cm}
A_d \sim m_{3/2}~y'_b 
\left(
\begin{array}{rrc}
0 & 5\,h~\varepsilon^3_d & -5\,k~\varepsilon^3_d \\
5\,h~\varepsilon^3_d & 7\,l~\varepsilon_d^2 & -7\,k^\prime~\varepsilon_d^2 \\
-5\,k~\varepsilon^3_d & -7\,k^\prime~\varepsilon_d^2 & 5
\end{array} 
\right).
\eeq

Going to the SCKM basis, after canonical normalization of the fields the trilinear matrices are given by

\begin{eqnarray} \nonumber
A_u & \longrightarrow & U_u.A_u.V_u^\dagger \, \sim \, m_{3/2}\,y_t\,
\left(
\begin{array}{rcc}
0.3\,\varepsilon_u^2\,\varepsilon_d^2 & ~~-0.7\,\varepsilon^2_u\,\varepsilon_d~~ & 0.3\,\varepsilon_u^2\,\varepsilon_d\\
-0.7\,\varepsilon^2_u\,\varepsilon_d & -7.6~\varepsilon_u^2 & -\,\varepsilon_u^2 \\
0.4\,\varepsilon^2_u\,\varepsilon_d & ~~~1.3~\varepsilon_u^2 & ~5.0
\end{array} 
\right), \\
\nonumber \\
\nonumber \\
A_d & \longrightarrow & U_d.A_d.V_d^\dagger ~ \sim ~ m_{3/2}~y_b~ 
\left(
\begin{array}{rrc}
-5.3~\varepsilon_d^4 & -2.3~\varepsilon^3_d & ~~2.3~\varepsilon^3_d \\
-2.3~\varepsilon^3_d & 5.2~\varepsilon_d^2 & -1.5~\varepsilon_d^2 \\
3.1~\varepsilon^3_d & -2.0~\varepsilon_d^2 & 5.0
\end{array} 
\right).
\end{eqnarray}

For the soft masses, the structure of the K\"ahler potential in the effective theory is given by \eq{kahlerab}, where as before, $b_{ij}$ encodes the flavor effects in the soft masses while $c_{ij}$ gives the form of the kinetic terms. Under our assumption that the $SU(2)_L$ doublet-messengers are much heavier than their singlet counterparts, their effects in the soft masses of the left-handed SM fields will be negligible; we therefore focus on the right-handed fields. 

The leading supergraphs can be found in Fig.~\ref{fig:fig03}. The first row of Fig.~\ref{fig:fig03} gives the dominant diagrams which contribute to the coefficients $c_{ij}$, while the second gives a subleading diagram which is supressed by $\alpha^2$. In addition, these diagrams will always have $\alpha^{2n}$ corrections resulting from inserting pairs of $\bar\phi_3$ fields with the flavor indices contracted internally. These corrections can be factored out, amounting to a global rescaling of the fields that plays no role here. Taking all this into account, we obtain, 

\beq
c^u_{ij} ~ \simeq ~
\left(\begin{array}{ccc}
\varepsilon_u^2\,\varepsilon_d^2 & -2\varepsilon_u^3 & 2\varepsilon_u^3 \\
-2 \varepsilon_u^3 & \varepsilon_u^2 & -(1 + \alpha^2) \varepsilon^2_u \\
2 \varepsilon_u^3 & -(1 + \alpha^2) \varepsilon^2_u & \alpha^2
\end{array} \right),~~
 c^d_{ij} ~ \simeq ~
\left(\begin{array}{ccc}
\varepsilon_d^4 & \varepsilon^3_d & -\varepsilon^3_d \\
\varepsilon^3_d & \varepsilon_d^2 & -(1 + \alpha^2) \varepsilon_d^2 \\
-\varepsilon^3_d & -(1 + \alpha^2)\varepsilon_d^2 & \alpha^2 \end{array} \right).
\eeq

Now, inserting the $XX^{\dagger}$ vertex in every possible position within the previous diagrams we can calculate the coefficients $b_{ij}$, which are found to be

\beq
b^u_{ij} ~ \simeq ~ \left(\begin{array}{ccc}
2 \varepsilon_u^2\,\varepsilon_d & -6 \varepsilon^3_u & 6 \varepsilon^3_u \\
-6 \varepsilon^3_u & 2 \varepsilon_u^2 & -(2 + 5\alpha^2)\varepsilon_u^2 \\
6\varepsilon^3_u & -(2 + 5\alpha^2)\varepsilon_u^2 &  2 \alpha^2
\end{array} \right) , ~~
b^d_{ij} ~ \simeq ~ 
\left(\begin{array}{ccc}
2\varepsilon_d^4 & 3\varepsilon^3_d & -3\varepsilon^3_d \\
3\varepsilon^3_d & 2\varepsilon_d^2 & -(2 + 5 \alpha^2)\varepsilon_d^2 \\
-3\varepsilon^3_d & -(2 + 5\alpha^2)\varepsilon_d^2 & 2 \alpha^2 \end{array} \right).
\eeq

As before, knowing $b_{ij}$, the soft mass matrices are given by $(m_{\mathrm{soft}}^2)_{ij}=m_{3/2}^2(\delta_{ij}+b_{ij})$. Then, the fields should be redefined first in the canonical basis and then in the SCKM basis. Finally, after these field redefinitions, the expression for the soft mass matrices for $\bar{u}$ and $ \bar{d}$ will be, 

\begin{eqnarray}
\left(m^2_{\mathrm{soft}}\right)_{\bar u}  &\simeq & m_{3/2}^2 \begin{pmatrix}
1+0.1\,\varepsilon_d^2 & -4~\varepsilon^3_u & ~~~3~\varepsilon^3_u \\
-4~\varepsilon^3_u & ~~1+\varepsilon_u^2~ & ~-3~\varepsilon_u^2~ \\
~~3~\varepsilon^3_u & -3~\varepsilon_u^2 & ~~1.3
\end{pmatrix}  \,, \nonumber \\
\left(m^2_{\mathrm{soft}}\right)_{\bar d} & \simeq & m_{3/2}^2 \begin{pmatrix}
1+2~\varepsilon_d^4 & 0.5~\varepsilon^3_d & ~~~~2~\varepsilon^3_d \\
~~~0.5~\varepsilon^3_d & ~1+3~\varepsilon^2_d~ & ~-2~\varepsilon_d^2 \\
-2~\varepsilon^3_d & -2~\varepsilon_d^2 & ~~~1.3
\end{pmatrix}\,.
\end{eqnarray}


\section{Flavor Phenomenology} \label{phenomenology}

In the previous sections, we obtained the soft-breaking matrices in two flavor symmetric models by simply assuming they were generated by the same mechanism responsible for generating the Yukawa couplings. Although these structures are to some extent dependent on the mediator sector, we have seen that integrating out the heavy mediator fields always generates non-universalities in the soft terms. In these flavor models, the structures of the soft matrices are fixed by  symmetry, no longer unknowns of the (124-)SUSY model, but {\em predictions}. 

Models in this spirit provide a new point of view for the phenomenology of flavor-supersymmetric models. Effectively, we have only the four parameters of the usual constrained MSSM, with $m_0$, $m_{1/2}$, $\tan \beta$ and sgn($\mu$) serving as inputs, the order of magnitude of the others being fixed by the flavor symmetry.\footnote{Due to our ignorance of the gauge kinetic function at $M_{\rm Pl}$, we take $m_{1/2}$ to be independent from $m_0 = m_{3/2}$.} Our aim is therefore not to constrain the flavor violating entries as functions of the sfermion masses, but rather to constrain the sfermion masses of the model using flavor observables, checking whether these constraints are competitive with direct LHC searches.  

To this end, we assume that the flavor symmetry is broken at a high scale of order $M_{\rm GUT}$, where the effective soft-breaking matrices are generated. These matrices must then be run using the MSSM renormalization group equations (RGE), in order to obtain the soft-breaking matrices at the electroweak scale and compared with the experimental constraints at low energies. The main effects of the RGE evolution are large contributions of order $\sim -2 m_{1/2}$ to the diagonal elements in the trilinear matrices of the first two generations, with slightly smaller contributions to the third generation (stop) entries. The diagonal elements of the soft-mass matrices receive a large gluino contribution of order $\sim 6 m_{1/2}^2$. On the other hand, off-diagonal entries in these matrices mostly remain unchanged by the running (possible exceptions are entries with $\order(1)$ Yukawa couplings)\cite{Falck:1985aa,Bertolini:1990if,Martin:1993zk,Yamada:1994id}. 

As an example of these ideas, we may apply the constraints provided by flavor observables to the toy $U(1)_f$ model of Section~\ref{sec:toy}. In this model, all squark soft mass matrices, $m_{\tilde Q}^2$,  $m_{\tilde{\bar{u}}}^2$ and  $m_{\tilde{\bar d}}^2$ have similar non-universal structures given by \eq{msoftU1}, while the trilinear couplings, $A_u$ and $A_d$ also have sizeable off-diagonal entries as displayed in \eq{trilinU1}. After RGE evolution, the corresponding matrices at the electroweak scale can be extracted, and the off-diagonal elements compared with the usual Mass Insertion (MI) bounds \cite{Gabbiani:1988rb,Hagelin:1992tc,Gabbiani:1996hi,Ciuchini:1998ix}. Taking a fixed squark mass scale, \eq{trilinU1} and \eq{msoftU1} give $(\delta^d_{12,13})_{\rm LL, RR} \simeq 2 \varepsilon^3/3\simeq 7.6 \times 10^{-3}$, $(\delta^d_{23})_{\rm LL, RR}\simeq 2 \varepsilon^2/3\simeq 3.4 \times 10^{-2}$, to be compared with the bounds, $(\delta^d_{12,13})_{\rm LL, RR}$ and $(\delta^d_{12,13})_{\rm LL, RR} \leq \cdots$ 

In case of LR transitions, our $U(1)$ model has a large right-handed mixing in the Yukawa matrices and therefore the largest mass insertions in this model are 
\begin{eqnarray} \nonumber
& &(\delta^d_{31})_{\rm LR} \simeq 2 \varepsilon/6  m_b \tan \beta/ m_{3/2} \simeq 0.0225 (50/\tan \beta) (500 {\rm GeV }/m_{3/2}) \\ \nonumber
& &(\delta^d_{32})_{\rm LL, RR}\simeq 0.05  (50/\tan \beta) (500 {\rm GeV }/m_{3/2}). 
\end{eqnarray}
The relevant bounds in this case are  $(\delta^d_{23,32})_{\rm LR} \leq 2 \times 10^{-2} (6 m_{3/2}^2/(500 {\rm GeV })^2)$, which, as we see, could be interesting in the case of large $\tan \beta$ and relatively low $m_{3/2}$ (taking into account the present LHC constraints on squark masses) and will be considered in a future, more phenomenological, work.

With these constraints, it is clear that the best observables for this model are the ones associated to transitions between the first and second generations, i.e.: $\Delta m_K$ and, if we take into account possible $\mathcal{O}(1)$ phases in these matrices, $\varepsilon_K$ and $\varepsilon^\prime/\varepsilon$  \cite{Masiero:2000rg}. To show the power of flavor observables in this $U(1)_f$ model, we will simply take the constraints from  $\Delta m_K$.\footnote{A full analysis using all available flavor observables is to appear in a future publication \cite{Das:2017unp}.}

\begin{figure}[htbp!]
\begin{minipage}{0.46\textwidth}
\centerline{\includegraphics[width=8cm,height=8cm]{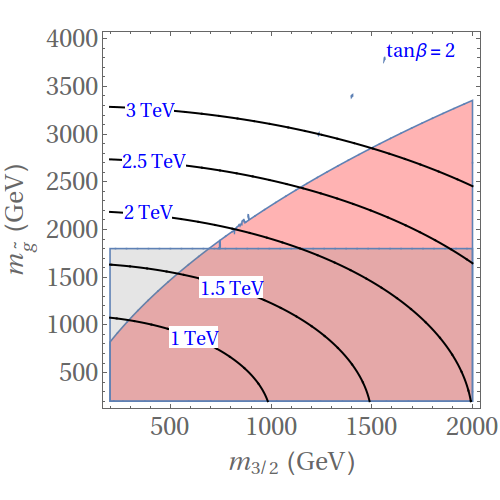}}
\caption{$\Delta m_K$ exclusion plot for the toy $U(1)$ model of the previous section. Points in red (darker shade) exceed the current experimental bounds. The rectangle in gray (light rectangular shade) represents the direct bound on gluino masses from LHC searches and the black contours correspond to different values of the average squark masses. Although we have taken $\tan \beta =2$ in this figure, the constraints are largely independent of $\tan \beta$.}
\label{fig:fig04}
\end{minipage}
\hfill
\begin{minipage}{0.46\textwidth}
\centerline{\includegraphics[width=8cm,height=8cm]{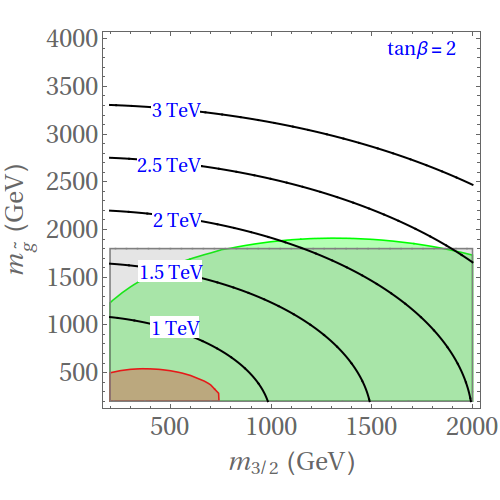}}
\caption{$\Delta m_K$ (red/dark gray) and $\varepsilon_K$ (green/light gray) exclusion plot for the $SU(3)_f$ model. Points in red (dark gray) or green (light gray) exceed the current experimental bounds. The rectangle in gray (shaded) represents the direct bound on gluino masses from LHC searches and the black contours correspond to different values of the average squark masses. Although we have taken $\tan \beta =2$ in this figure, the constraints are largely independent of $\tan \beta$.}
\label{fig:fig05}
\end{minipage}
\end{figure}

Although flavor symmetries fix the orders of magnitude of different entries in the soft-breaking matrices in powers of $\varepsilon$, unless we completely define the mediator sector, we cannot predict the $\order(1)$ coefficients. In particular, as we do not know the sign (phase) of these entries, both constructive or destructive interference with the SM contribution are possible. Taking a conservative approach following \cite{Gabbiani:1996hi}, we require that any supersymmetric contribution, (including both different mass insertions and additional supersymmetric particles such as the gluino and chargino), to a given observable must be smaller than the measured experimental value. In this way, we obtain the results given in Fig.~\ref{fig:fig04}. Here, we consider only the dominant contribution given by the gluino, with $(\delta_{12})_{\rm LL}\times (\delta_{12})_{\rm RR}$ interference. 

The region in red (darker shade) is excluded from $\Delta m_K$, while the region in grey (lighter rectangular shade) is ruled out from direct gluino searches \cite{Aad:2016jxj,Aad:2016qqk,Aad:2016eki,Aaboud:2016tnv,Aaboud:2016zdn,Khachatryan:2016kdk}. The contours in black give an estimate for the average squark masses, obtained from the approximate one-loop RGE relation $m_{\tilde q}^2 \sim  6 m_{1/2}^2 + m_{3/2}^2$. As can be seen from the figure, the experimental bounds from $\Delta m_K$ can be competitive with direct searches for a large range of gluino and gravitino masses for this simple $U(1)$ flavor symmetry. Taken in conjunction with the direct gluino bounds, the flavor constraints in the $U(1)_f$ model can also push up the lower bound for the gluino and squark masses as a function of $m_{3/2}$. The fact that the bounds are stronger for larger $m_{3/2}$ is simply due to the flavor-changing entries being proportional to $m_{3/2}$ and having taken the gluino mass independent from $m_{3/2}$. For the particular case of $m_{1/2}=m_{3/2}$ at $M_{\rm Pl}$, we have $m_{\tilde g}\simeq 2.7 m_{3/2}$ and thus flavor bounds would require $m_{\tilde g}, m_{\tilde q} \gtrsim 2.1$ TeV, improving significantly bounds from direct searches in the $U(1)_f$ scenario. These bounds would be much stronger if these off-diagonal entries had sizeable imaginary parts and we were to apply the limits from $\varepsilon_K$.

We can apply the same procedure to the $SU(3)_f$ model. In this case, as we have taken the left-handed mediator masses $M_L \gg M_u,M_d$, we can neglect the flavor non-universality in $m_{\tilde Q}^2$, considering only  $m_{\tilde{\bar{d}}}^2$ and  $m_{\tilde{\bar u}}^2$ apart from the trilinear couplings $A_u$ and $A_d$.
Using the same reasoning as for the $U(1)$ model, the most sensitive observables are still $\Delta m_K$ and $\varepsilon_K$, although the bounds from  $\Delta m_K$ are much weaker due to the absence of the left-handed mass insertions, which in the $U(1)$ model give the largest contribution to $\Delta m_K$ through a mixed term $(\delta_{12})_{\rm LL} (\delta_{12})_{\rm RR}$. The dominant contribution is now given by the gluino with $(\delta_{12})_{\rm LR}^2 + (\delta_{12})_{\rm RL}^2$ mass insertions. In this case, $\Delta m_K$ would exclude only a ``small'' region with $m_{\tilde g} \simeq m_{\tilde q} \lesssim 500$ GeV. However, if we assume that these flavor-changing entries have imaginary parts of $\order(1)$\cite{Botella:2004ks,Botella:1994cs,Santamaria:1993ah}, \footnote{Sizeable phases in these mass matrices are found in models of spontaneous CP violation in the flavor sector \cite{Ross:2004qn,Calibbi:2008qt,Calibbi:2009ja} which naturally explain the presence of phases in the CKM matrix. However, due to re-phasing freedoms, the imaginary parts are typically subleading.} we can apply the limits from $\varepsilon_K$. In this case, the bounds are much stronger and flavor limits can compete with direct LHC searches for the $SU(3)$ model as well. As we see, the $SU(3)_f$ model is safer than the $U(1)_f$ model, both due to the non-Abelian nature of the symmetry, which relates different generations, and to the small number of flavon insertions in the $SU(3)_f$ model, allowing flavor off-diagonal entries only at higher orders.   

Although these bounds are model (and mediator) dependent, similar searches, using a complete set of flavor observables, may be applied to any supersymmetric flavor model. A more complete analysis of these effects for different flavor symmetries will be presented in a future work \cite{Das:2017unp}. As we have shown for these simple models, these phenomenological studies may provide a very important tool in ruling out or discovering potential flavor symmetries.


\section{Conclusions} \label{conclusions}

The peculiar structure of the flavor sector remains one of the most puzzling and intriguing legacies of the SM. The question of whether these patterns arise from a deeper underlying theory, pointing the way to new physics, remains to be answered. Although the hope is that this will indeed be the case, it is clear that additional experimental inputs in the form of new observable flavor interactions would provide a clear boon towards unravelling this mystery.

New flavor interactions inherent in supersymmetric theories can provide an excellent laboratory in which to probe models where this underlying theory is governed by a flavor symmetry. Under our assumptions that the soft-breaking terms respect the flavor symmetry and that SUSY breaking is primarily communicated to the visible fields by gravity, these new flavor structures are calculable and governed by the flavor symmetry and the mediator sector of the underlying theory. The large number of parameters in the MSSM is thus vastly reduced.

The main result of this work is that even when the ultraviolet theory is flavor symmetric, the soft-terms of the effective theory  below $\Lambda_f$ can be strongly non-universal. We have demonstrated this explicitly in the case of two specific models: one with an Abelian $U(1)$ symmetry, the other an $SU(3)$. 

As these effects are calculable, their phenomenological implications for flavor observables can in principle provide strong constraints on the parameter space of these models.  In some cases, these constraints may be competitive with direct LHC searches. With the soft matrices given, indirect flavor measurements can provide a new tool to provide constraints on parameters like the gluino mass or the average scale of the squark masses. 

Unfortunately, these effects are model dependent, and an obvious extension of this work would be an application of these techniques to a representative set of popular flavor models available in the literature. This would require a more complete phenomenological analysis, taking into account all available flavor observables. In addition, although we have not specified the scale of flavor breaking, if the flavor symmetry is gauged and the scale of its breaking sufficiently light, one may also superimpose the constraints from the associated $Z^\prime$ boson, leading to a further restriction of the parameter space. We have also chosen to focus only on the quark sector, and it would be interesting to see whether these techniques can provide interesting constraints in lepton flavor models. 

In highlighting these new sources of flavor violation, we hope to provide a new tool for tackling the flavor puzzle in one of the most well studied frameworks for new physics, SUSY. 

\section*{Acknowledgments}
The authors are grateful to A. Santamaria for useful discussions. This work has been partially supported under MEC and MINECO
Grants FPA2011-23596, FPA2011-23897 and FPA2014-57816-P and  by the ``Centro de Excelencia Severo Ochoa'' Programme under
grant SEV-2014-0398. OV acknowledges partial support from the ``Generalitat Valenciana'' grant  PROMETEOII/2013/017. All Feynman
diagrams have been drawn using Jaxodraw\cite{Binosi:2003yf,Binosi:2008ig}.


\bibliographystyle{JHEP}
\bibliography{references.bib}
\end{document}